\documentclass[superscriptaddress,twocolumn]{revtex4}
\usepackage{graphicx,epsfig}
\usepackage{amsmath}
\usepackage {amssymb}

\newcommand{\be}{\begin{eqnarray}}
\newcommand{\ee}{\end{eqnarray}}
\newcommand{\bea}{\begin{eqnarray}}

\newcommand{\eea}{\end{eqnarray}}

\def\k{\kappa}

\makeindex

\begin{document}

\title{Scalar quasinormal modes for $2+1$-dimensional Coulomb like AdS black holes from non lineal electrodynamics}
\author{Almendra Arag\'{o}n}
\email{almendra.aragon@mail.udp.cl} \affiliation{Facultad de
Ingenier\'{i}a y Ciencias, Universidad Diego Portales, Avenida Ej\'{e}rcito
Libertador 441, Casilla 298-V, Santiago, Chile.}
\author{P. A. Gonz\'{a}lez}
\email{pablo.gonzalez@udp.cl} \affiliation{Facultad de
Ingenier\'{i}a y Ciencias, Universidad Diego Portales, Avenida Ej\'{e}rcito
Libertador 441, Casilla 298-V, Santiago, Chile.}
\author{Joel Saavedra}
\email{joel.saavedra@ucv.cl} \affiliation{Instituto de
F\'{i}sica, Pontificia Universidad Cat\'olica de Valpara\'{i}so,
Casilla 4950, Valpara\'{i}so, Chile.}
\author{Yerko V\'{a}squez}
\email{yvasquez@userena.cl}
\affiliation{Departamento de F\'{\i}sica, Facultad de Ciencias, Universidad de La Serena,\\
Avenida Cisternas 1200, La Serena, Chile.}

\date{\today}

\begin{abstract}

We study the propagation of scalar fields in the background of $2+1$-dimensional Coulomb like AdS black holes, and we show that such propagation is stable under Dirichlet boundary conditions. Then, we solve the Klein-Gordon equation by using the pseudospectral Chevyshev method, and we find the quasinormal frequencies. Mainly, we find that the quasinormal frequencies are purely imaginary for a null angular number and they are complex and purely imaginary for a non null value of the angular number, which depend on the black hole charge, angular number and overtone number. On the other hand, the effect of the inclusion of a Coulomb like field from non lineal electrodynamics to General Relativity for a vanishing angular number is the emergence of two branches of quasinormal frequencies in contrast with the static BTZ black hole.    

\end{abstract}

\maketitle


\tableofcontents


\section{Introduction}

Three-dimensional models of gravity have been of great interest due to their simplicity over four-dimensional and higher-dimensional models of gravity, and since some of their properties are shared by their higher dimensional analogs. In this sense, the Bañados, Teitelboim, and Zanelli (BTZ) black hole \cite{Banados:1992wn} shares several features of Kerr black holes \cite{Carlip:1995qv}. The electrically charged BTZ black hole was studied in Ref. \cite{Banados:1992wn},  and the charged and rotating one was presented in Ref. \cite{Martinez:1999qi}. In both cases the dynamics of the gauge field was defined by the usual Maxwell Lagrangian, and consequently, the gauge field exhibits a logarithmic dependence on the radial coordinate, and it was shown that the electrically charged black hole is pathological due to it exists for arbitrarily negative values of the mass, and there is no upper bound on the electric charge \cite{Martinez:1999qi}. However, the usual Maxwell action is not invariant under conformal transformations of the metric. In this way, by introducing a nonlinear electrodynamics as the source of the Einstein equation, it is possible to have an action invariant under conformal transformations, which was studied in Ref. \cite{Cataldo:2000we}, and exact black hole solutions that describe spacetimes of constant curvature with a genuine singularity at the origin were found. In this case, the electric field has the Coulomb form of a point charge in the Minkowski spacetime, and the solutions describe charged (anti)–de Sitter spacetimes, and there is not a logarithmic function in the metric expression \cite{Cataldo:2000we}. Their thermodynamic properties through both the standard and geometrothermodynamics approaches were recently studied in Ref. \cite{Cataldo:2020cxm}. Actually, it is known that the invariance under conformal transformations is recovered in any spacetime dimension $n$ if the Maxwell Lagrangian is raised to 
the $(n/4)^{th}$ power, which provides a Coulomb-like electric field in arbitrary dimensions \cite{Hassaine:2007py}. Then, this kind of theories were extended in order to study the existence of hairy black holes solutions \cite{Cardenas:2014kaa}.  

In this work, we consider  $(2+1)$-dimensional Coulomb like AdS black holes as background, and we study the stability of the propagation of a scalar field in order to study the effects of the black hole charge on the propagation of neutral scalar fields, by using the pseudospectral Chebyshev method \cite{Boyd}, which is an effective method to find high overtone modes \cite{Finazzo:2016psx,Gonzalez:2018xrq,Becar:2019hwk,Aragon:2020qdc, Aragon:2020tvq, Aragon:2020xtm, Aragon:2020teq, Fontana:2020syy}. It is worth mentioning that nonlinear field theories are of interest to different branches of mathematical physics because most physical systems are inherently nonlinear in nature.  Also, exact solutions of charged black holes under different frameworks considering general relativity or modified gravity have been reported \cite{Dey:2004yt,Cai:2004eh,Zou:2013owa,Boillat:1970gw,Fernando:2003tz,Jing:2010zp,deOliveira:1994in,Gullu:2010pc,Hendi:2014mba,Hendi:2016pvx,Hendi:2017mgb,Hendi:2017oka,Hendi:2020yah}, and regular charged black holes \cite{AyonBeato:1998ub}. 

The stability of black holes and also the behaviour of the propagation of fields in black hole backgrounds has been studied for a long time starting from the pioneering work by Regge and Wheeler \cite{Regge:1957a}. Also, it has been found that most of black holes are stable under various types of perturbations, see \cite{Konoplya:2011qq}. In this context, an effective way to study the stability of black holes is to calculate the  quasinormal modes (QNMs) and their quasinormal frequencies (QNFs) \cite{Zerilli:1971wd,
Zerilli:1970se, Kokkotas:1999bd, Nollert:1999ji}. Also, the QNMs have recently acquired great interest due to the detection of gravitational waves \cite{Abbott:2016blz}, and the QNMs have a recognizable importance in the context of the correspondence AdS/CFT \cite{Maldacena:1997re, Aharony:1999ti,Horowitz:1999jd}. The behaviour of QNMs for minimally coupled field
has been studied in Refs. \cite{Horowitz:1999jd, Konoplya:2002zu,Cardoso:2003cj}. The QNMs of the BTZ black hole for conformal scalar field was studied in Ref. \cite{Chan:1996yk}, and for non-conformal scalar, electromagnetic and dirac fields in \cite{Cardoso:2001hn}. Also, the QNMs of the BTZ black hole surrounded by the conformal scalar field was analyzed in Ref. \cite{Konoplya:2004ik}, where it was estimated the shift in the quasinormal spectrum of the BTZ black hole stipulated by the back reaction of the Hawking radiation. See \cite{Gupta:2015uga, Gupta:2017lwk}, for the scalar, and fermionic quasinormal modes of the BTZ black hole in the presence of spacetime noncommutativity, respectively, and see  \cite{Becar:2014jia, Becar:2015kpa, Gonzalez:2015gla, Gonzalez:2017shu, Panotopoulos:2017hns, Gonzalez:2017zdz,Ciric:2017rnf, Rincon:2018sgd, Destounis:2018utr, Panotopoulos:2019qjk, Ciric:2019uab}, for other charged geometries.

This work is organized as follows. In Sec.~\ref{background} we give a brief review of three-dimensional Coulomb like AdS black holes. Then, in Sec.~\ref{QNM} we study the stability under Dirichlet boundary conditions and we calculate the QNFs of  scalar  perturbations numerically by using the pseudospectral Chevyshev method. Finally, we conclude in Sec.~\ref{conclusion}.

\section{Three-dimensional Coulomb like AdS black holes}
\label{background}

The action of the (2+1)-Einstein theory coupled with nonlinear electrodynamics is given by \cite{Cataldo:2000we} 
 \begin{eqnarray} \label{action}
 S=\int d^{3}x\sqrt{-g}\left(\frac{1}{16 \pi}\left(R-2 \Lambda\right)+L(F)\right)~,
 \end{eqnarray}
where $R$ is the Ricci scalar, $\Lambda$ corresponds to the cosmological constant, and $L(F)=C\left|  F\right| ^{3/4}$ represents the electromagnetic Lagrangian. The metric solution of this theory is given by 
\begin{eqnarray}
 ds^{2}=-f(r)dt^{2}+f^{-1}(r)dr^{2}+r^2 d \phi^2\,, \label{metricBH}
 \end{eqnarray}
where the lapse function is
\begin{equation}
 f(r)=-M-\Lambda r^2+\frac{4q^2}{3r}\,, \label{f(r)}
 \end{equation} 
where $M$ is a constant related to the physical mass, and $q$ is a constant related to physical charge. Also, the electric field is given by
\begin{equation}
 E(r)=\frac{q}{r^2}\,.
\end{equation}
The spacetime is asymptotically de-Sitter spacetime for $\Lambda > 0$, asymptotically flat for $\Lambda = 0$, and asymptotically anti de-Sitter for $\Lambda <0$. The roots of the lapse function are  
\begin{equation}
r_{h_{1}}=\frac{h}{3 \Lambda}-\frac{M}{h}~,\,\,\label{rh}
\end{equation}
\begin{equation}
r_{h_{2}}=-\frac{h}{6\Lambda}+\frac{M}{2h}+i\frac{\sqrt{3}}{2}\left(\frac{h}{3 \Lambda}+\frac{M}{h}\right)~,\,\,
\end{equation}
\begin{equation}
r_{h_{3}}=-\frac{h}{6\Lambda}+\frac{M}{2h}-i\frac{\sqrt{3}}{2}\left(\frac{h}{3 \Lambda}+\frac{M}{h}\right),
\end{equation}
where $h$ is 
\begin{equation}
h=\left(\left(18q^2+3\sqrt{3\left(\frac{M^3}{\Lambda}+12q^4\right)}\right)\Lambda^2\right)^{\frac{1}{3}}~.
\end{equation}

Here, we focus our study in the AdS case, where $M>0$, the solution shows  different behaviors for the geometry depending the value of the cosmological constant. There is a black hole solution with inner and outer horizons when $ 0>\Lambda >- \frac{M^3}{12q^4}$,
there is one real and two complex solutions when $ \Lambda <- \frac{M^3}{12q^4}$. Finally, when $\Lambda =- \frac{M^3}{12q^4}$, the solution represents an extreme black hole. \\

\section{Scalar perturbations }
\label{QNM}

\begin{figure}[h]
\begin{center}
\includegraphics[width=0.5\textwidth]{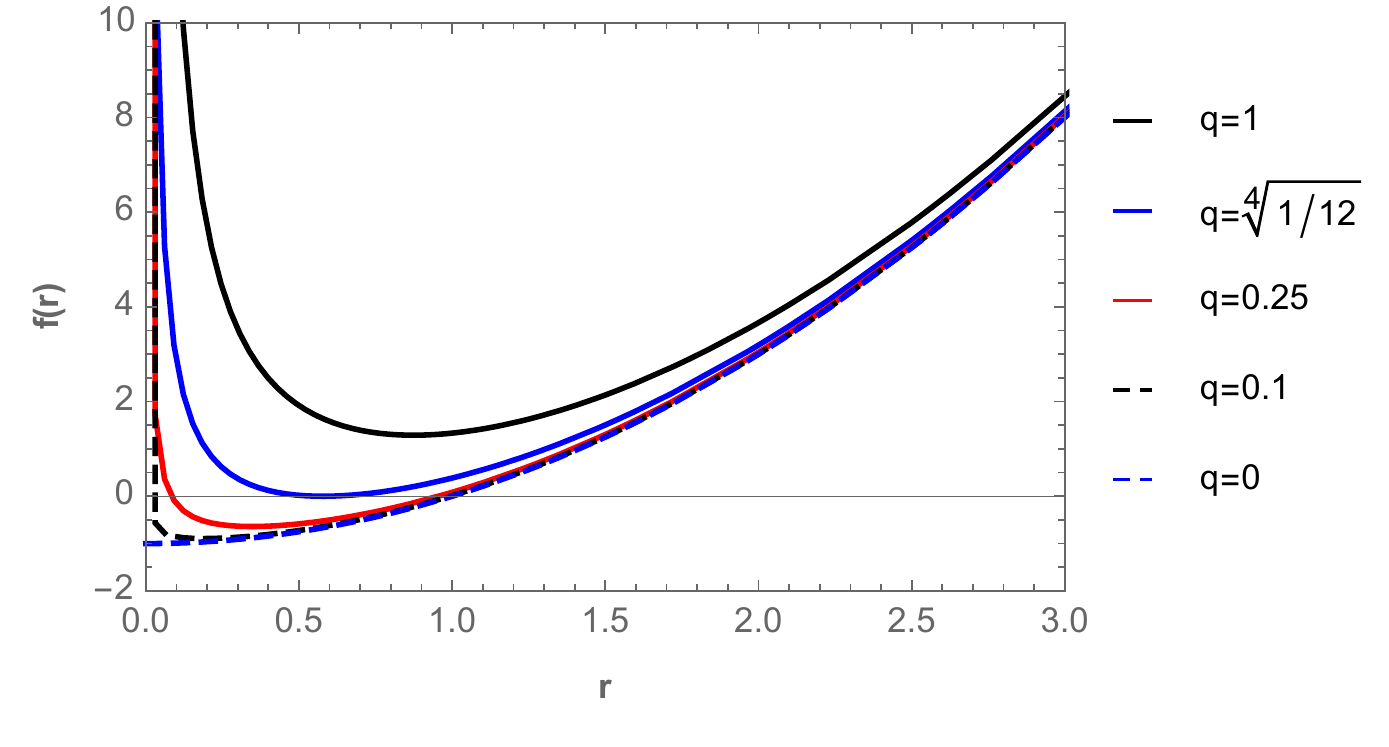}
\end{center}
\caption{The behaviour of $f(r)$ with $M=1$, $\Lambda=-1$, and different values of the charge $q$.}
\label{Lapse}
\end{figure}

\begin{figure}[h]
\begin{center}
\includegraphics[width=0.5\textwidth]{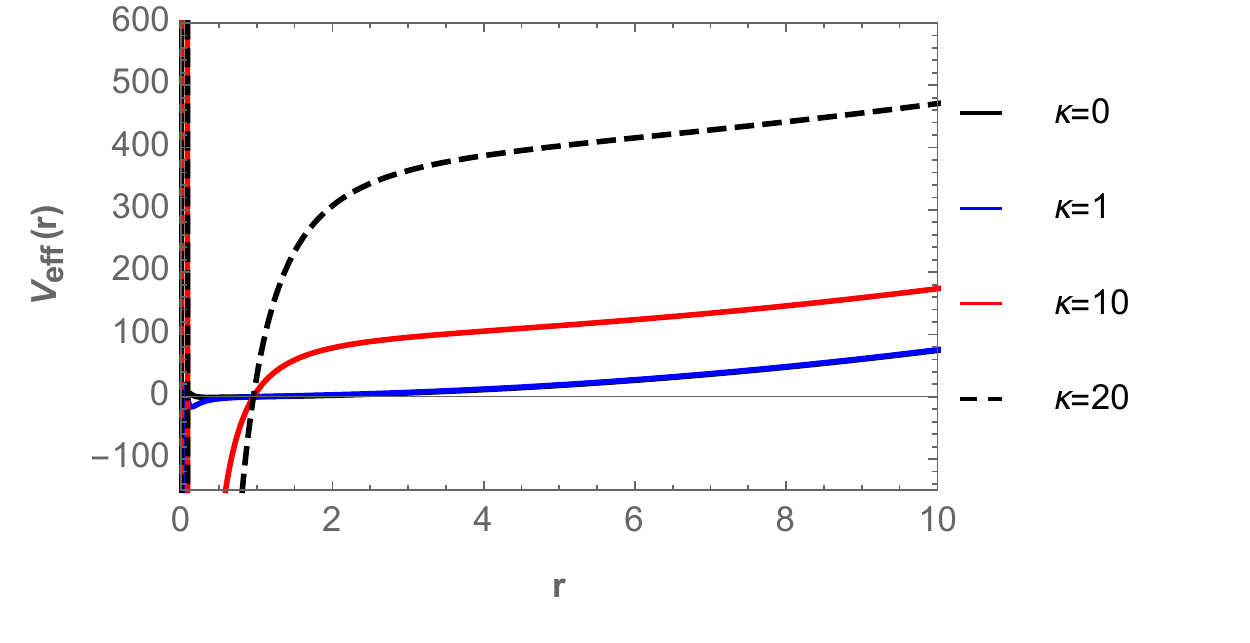}
\includegraphics[width=0.5\textwidth]{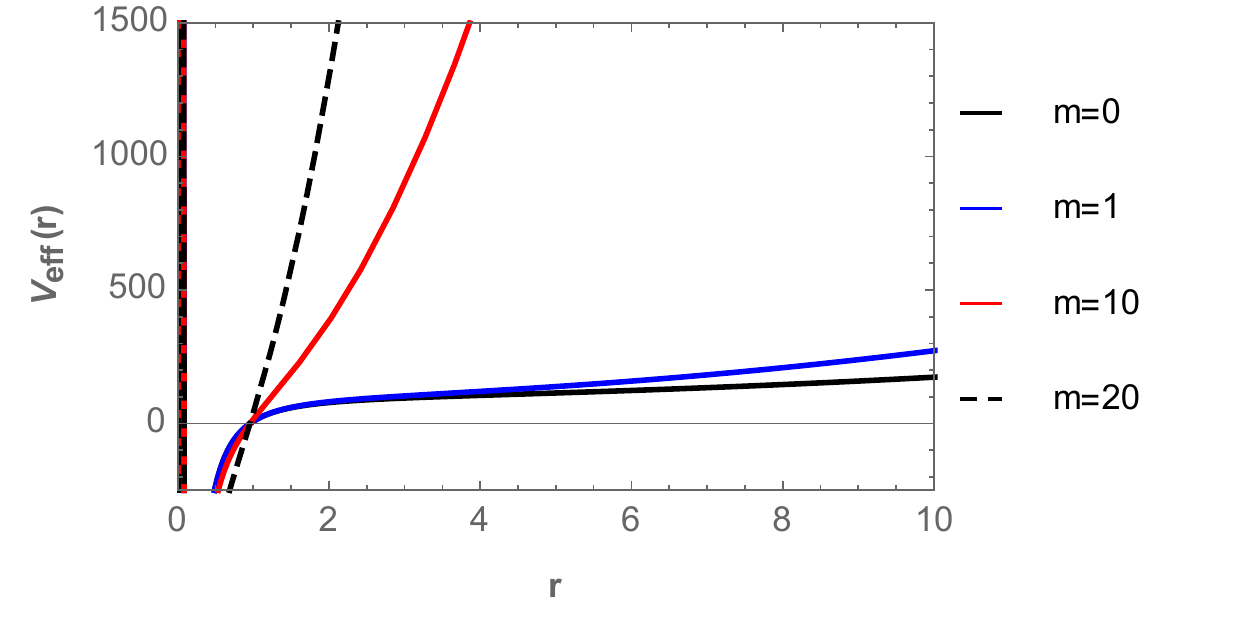}
\end{center}
\caption{The behaviour of $V_{eff}(r)$ with $M=1$, $\Lambda=-1$, and $q=0.25$. Top panel for massless scalar field and different values of the angular number $\kappa$, and bottom panel for a fix value of $\kappa=10$ and different values of the scalar field mass $m$. Here, $r_-\approx 0.084$, and $r_+\approx 0.955$.}
\label{Potential}
\end{figure}

In order to obtain the QNMs of scalar perturbations in the background of  the metric (\ref{f(r)})
we consider the Klein-Gordon equation
\begin{equation}
\label{KGE}
\Box \psi = \frac{1}{\sqrt{-g}}\partial_{\mu}\left(\sqrt{-g} g^{\mu\nu}\partial_{\nu}\right)\psi=m^2\psi~,
\end{equation}
with suitable boundary conditions for a black hole geometry. In the above expression $m$ is the mass of the scalar field $\psi $. The Klein-Gordon equation can be written as 
\begin{equation}
\frac{d}{dr}\left(r f(r)\frac{dR}{dr}\right)+\left(\frac{r\omega^2}{f(r)}-\frac{\kappa^2}{r}-m^{2}r \right) R(r)=0\,, \label{radial}
\end{equation}%
by means of the ansatz $\psi =e^{-i\omega t} e^{i\kappa \phi} R(r)$, where $\kappa=0, 1,2, \dots$. Then, defining $R(r)$ as $R(r)=\frac{F(r)}{\sqrt{r}}$, and by using the tortoise coordinate $r^*$ given by $dr^*=\frac{dr}{f(r)}$, the Klein-Gordon equation can be written as 
\begin{equation}
 \label{ggg}
 \frac{d^{2}F(r^*)}{dr^{*2}}-V_{eff}(r)F(r^*)=-\omega^{2}F(r^*)\,,
 \end{equation}
that corresponds to a one-dimensional Schr\"{o}dinger equation
 with an effective potential $V_{eff}(r)$, which is parametrically thought as $V_{eff}(r^*)$, and it is given by
 \begin{widetext}
  \begin{eqnarray}\label{pot}
 \nonumber V_{eff}(r)&=&\frac{f(r)}{r^2} \left(\kappa^2 + r\left( m^2r+ \frac{f^\prime(r)}{2}\right) -\frac{f(r)}{4}\right)~.\\
 &&-\frac{\left(3 r \left(M+\Lambda  r^2\right)-4 q^2\right) \left(r \left(4 \kappa ^2+r^2 \left(4 m^2-3 \Lambda \right)+M\right)-4 q^2\right)}{12 r^4}
 \end{eqnarray}
 \end{widetext}
In Fig. \ref{Lapse} we plot the lapse function, where for fixed values of the black hole mass and cosmological constant,  it is possible to observe the transition among a black hole with one horizon ($q=0$), a black hole with two horizons ($q=0.25$), the extremal one ($q_{ext}=(1/12)^{1/4}$) and a naked singularity ($q=1$). Notice that, for $q=0$, the spacetime is described by the static BTZ black hole. Then, in Fig. \ref{Potential} we show the effective potential, for fixed values of the cosmological constant, black hole mass and charge, in the top panel for different values of the angular number $\kappa$ and massless scalar field, and in the bottom  panel for a fixed value of the angular number and different values of scalar field mass. 
We can observe in the region outside the event horizon the potential is positive, and it increases when the angular number increases or when the scalar field mass increases. \\

\newpage

Also, we observe that 
there exist two different behaviours for the potential, one of them occurs for big values of the angular number where it is possible to observe a change in the concavity of the curve for  the potential, while the other is a monotonically increasing function, for small values of the angular number, see Fig. \ref{Potential} top panel. 
Also, it is possible to observe a similar behaviour depending on the scalar field mass, see Fig. \ref{Potential} bottom panel.
It is worth mentioning that it was shown that a potential-step type provides the purely imaginary QNFs, while the potential-barrier type gives the complex QNFs of a scalar field for the charged dilaton black hole \cite{Myung:2008pr}. Also, it was shown that the presence of the bump near the horizon explains clearly why the QNFs for gravitational and electromagnetic perturbations of the small SAdS black hole are complex in \cite{Cardoso:2001bb}. Moreover, for the four-dimensional Einstein-Gauss-Bonnet AdS black hole, it was shown that the existence of  purely imaginary or complex QNFs is related to the change of concavity of the potential at the event horizon \cite{Aragon:2020qdc}.
In this case, the second derivative of the effective potential evaluated at the event horizon $r_H$ is given by
 
  \begin{widetext}
 \begin{equation}
V_{eff}''(r_H) = \frac{16 q^2 r_H \left(6 \kappa ^2+m^2 r_H^2+6 M\right)-12 \Lambda  m^2 r_H^6-9 M r_H^2 \left(4 \kappa ^2+M\right)-160 q^4+9 \Lambda ^2 r_H^6}{6 r_H^6}\,.
 \end{equation}
 \end{widetext}
 
Expression that we will use in order to analyze the behaviour of the QNFs.

\subsection{Scalar field stability}

Now, in order to study the stability of the propagation of scalar fields in the background of three-dimensional Coulomb like AdS black hole, we follow the general argument given in Ref. \cite{Horowitz:1999jd}.
Thus, by replacing $\psi(r)=e^{i\omega r^{\ast}} F(r)$, in the Schr\"odinger like equation \eqref{ggg} we obtain
\begin{equation}\label{KleinFink}
\frac{d}{dr}(f(r)\frac{d\psi(r)}{dr})-2i\omega \frac{d\psi(r)}{dr}-\frac{V_{eff}(r)}{f(r)}\psi(r)=0\,.
\end{equation}
Then, multiplying Eq. (\ref{KleinFink}) by $\psi^{\ast}$ and performing integrations by parts, where we have considered Dirichlet boundary condition for the scalar field at spatial infinity, it is possible to obtain the following expression
\begin{equation}\label{relacion}
\int _{r_{+}}^{\infty}dr \left( f(r) \left|  \frac{d\psi}{dr}\right|^2+\frac{V_{eff}(r)}{f(r)} \left| \psi \right| ^2  \right)=-\frac{\left|\omega \right|^2 \left| \psi (r=r_{h})\right| ^2}{Im(\omega)}\,.
\end{equation}
So, notice that for scalar perturbations the potential (\ref{pot}) is positive outside the horizon; thereby the left hand side of \eqref{relacion} is strictly positive, which demand that $Im (\omega)<0$. Thus, it is possible to conclude that the propagation of scalar field respecting Dirichlet boundary conditions is stable. In the next section, we will carry out a numerical analysis in order to study the behaviour of the QNFs on the parameters that describe the scalar field and the black hole charge.

\subsection{Numerical Analysis}

There are several numerical methods in order to obtain the QNFs, some of them are the Mashhoon method, Chandrasekhar-Detweiler method, Wentzel-Kramers-Brillouin(WKB) method, Frobenius method, method of continued fractions, asymptotic iteration method (AIM) and improved AIM. For an extensive review, see \cite{Konoplya:2011qq}. Here, in order to solve numerically the differential equation (\ref{radial}) we consider the pseudospectral Chebyshev method \cite{Boyd}. Firstly, it is convenient to perform the change of variable  $y=1-r_H/r$. Thus, 
the values of the radial coordinate are limited to the range $[0,1]$, and the radial equation (\ref{radial}) becomes
\begin{widetext}
\begin{equation} \label{r}
(1-y)^3 f(y) R''(y) +\left( (1-y)^3 f'(y)-(1-y)^2 f(y) \right) R'(y) + \left( \frac{\omega^2 r_H^2}{f(y) (1-y)}- \kappa^2 (1-y) -\frac{m^2 r_H^2}{1-y} \right) R(y)=0\, ,
\end{equation}
\end{widetext}
where the prime denotes derivative with respect to $y$. Now, the event horizon is located at $y=0$ and the spatial infinity at $y=1$. So, in order to propose an ansatz, we analyze the behaviour of the differential equation at the horizon, and at infinity. In the neighborhood of the horizon the function $R(y)$ behaves as 
\begin{equation}
R(y)=C_1 e^{-\frac{i \omega r_H}{f'(0)} \ln{y}}+C_2 e^{\frac{i \omega r_H}{f'(0)} \ln{y}} \,,
\end{equation}
where the first term represents an ingoing wave and the second represents an outgoing wave near the black hole horizon. So, imposing the requirement of only ingoing waves on the horizon, we fix $C_2=0$. On the other hand, at infinity the function $R(y)$ behaves as
\begin{equation}
R(y)= D_1 (1-y)^{1 + \sqrt{1 -\frac{ m^2}{\Lambda}}}+ D_2 (1-y)^{1 - \sqrt{1 -\frac{ m^2}{\Lambda}}} \,.
\end{equation}
So, imposing that the scalar field vanishes at infinity requires $D_2=0$. Therefore, an ansatz for $R(y)$ is $(1-y)^{1 + \sqrt{1 -\frac{ m^2}{\Lambda}}}e^{-\frac{i \omega r_H}{f'(0)} \ln{y}} F(y)$, and by inserting it in Eq. (\ref{r}), it is possible to obtain a differential equation for the function $F(y)$. Now, to use the pseudospectral method, $F(y)$ must be expanded in a complete basis of functions $\varphi_i(y)$: $F(y)=\sum_{i=0}^{\infty} c_i \varphi_i(y)$, where $c_i$ are the coefficients of the expansion, and we choose the Chebyshev polynomials for the complete basis of functions, which are defined by $T_j(x)= \cos (j \cos^{-1}x)$, where $j$ corresponds to the grade of the polynomial. The sum must be truncated until some $N$ value, therefore the function $F(y)$ can be approximate by  
\begin{equation}
F(y) \approx \sum_{i=0}^N c_i T_i (x)\,.
\end{equation}
Thus, the solution is assumed to be a finite linear combination of the Chebyshev polynomials,
that are well defined in the interval $[-1,1]$. Due to the variable $y$ is defined in the interval $[0,1]$, $x$ and $y$ are related by $x=2y-1$.

Then, the interval $[0,1]$ is discretized at the Chebyshev collocation points $y_j$ by using the so-called Gauss-Lobatto grid where
\begin{equation}
    y_j=\frac{1}{2}[1-\cos(\frac{j \pi}{N})]\,, \,\,\,\, j=0,1,...,N \,.
\end{equation}
After that, the differential equation is evaluated at each collocation point. So, a system of $N+1$ algebraic equations is obtained, which corresponds to a generalized eigenvalue problem and it can be solved numerically to obtain the QNMs spectrum, by employing the built-in Eigensystem[ ] procedure  in Wolfram’s Mathematica \cite{WM}.\\

In this work, we use a value of $N$ into the interval [80-120] for the majority of the cases with an average running time in the range [50s-140s] which depends on the convergence of $\omega$ to the desired accuracy. For some especial cases, near extremal cases, we use a value of $N$ into the interval [200-240] with an average running time in the range [920s-1620s]. We use an accuracy of eight decimal places. 
Also, in order to guarantee the accuracy of the results the code was executed for several increasing values of $N$ until no difference was observed in the value of the QNF. Also, the complete parameter space associated to the models is $M\geq 0$, $\Lambda<0$, and $\kappa= \pm 1, \pm 2, ...$. Here, the regions of the parameter space explored is $M=1$, $\Lambda=-1$, and $0 \leq q \leq 0.536$, due to in this region is guaranteed the existence of two positive real roots for the lapse function, and a discrete set of values of $\kappa$ in the interval [0, 30]. Also, for the scalar field mass we consider a discrete set of values in the interval [0, 0.3]. \\

It is worth to mention that for uncharged black hole ($q=0$), the metric reduces to the static BTZ metric, and the QNMs in this case are given by \cite{Cardoso:2001hn,Birmingham:2001pj}
\begin{equation}
\label{BTZ}
    \omega=\pm\frac{\kappa}{\ell}-\frac{2\sqrt{M}(n+1)i}{\ell}\,.
\end{equation}

Also, it was shown that in the scalar and fermionic cases the vanishing boundary conditions at infinity are automatically satisfied for the exact solutions, which implies a spectrum of QNFs without a decay rate for the extremal rotating BTZ black hole \cite{Crisostomo:2004hj}. Also, Dirac quasinormal modes of rotating BTZ black holes with torsion was studied in Ref. \cite{Becar:2013qba}.

\subsubsection{Massless scalar fields}

In order to analyze the stability of the propagation of massless scalar fields we consider a null and non null angular number separately.\\

\begin{itemize}

\item{\bf{Case $\kappa = 0$}}. The behaviour of the QNFs for massless scalar fields 
for fixed values of black hole mass, cosmological constant, and a vanishing angular number, and different values of the overtone number, and black hole charge, is shown in Fig. \ref{Coulomb1}. Here, by comparing with the QNFs for the static BTZ black hole (\ref{BTZ}), it is possible to observe that for a vanishing angular number the frequencies remain being purely imaginary. Also, note that the inclusion of the charge implies the emergence of two branches, while for one of them the decay rate always decreases, and it is less than the decay rate for the static BTZ background, the other one has a sector where the decay rate is bigger than the decay rate for the static BTZ background, and it increases when the black hole charge increases until a maximum value, then it decreases, and then there is a sector where the decay rate is less than the decay rate for the static BTZ metric. On the other hand, the separation of each branch for a fixed value of the black hole charge increases when the black hole charge increases, until a maximum value, and then this separation begin to decreases and the black hole becomes in a near-extremal one. The numerical values for $n=0,1,2,3$ are in Table \ref{TCoulomb1}. On the other hand, for $q=0$, $M=1$, $\Lambda=-1$, and $\kappa=0$, the event horizon is located at $r_H=1$, and the second derivative of the potential is null.
However, where we consider a not null charge, the second derivative of the potential is 
\begin{equation}
\frac{-160 q^4+96 q^2 r_H+9 r_H^2 \left(r_H^4-1\right)}{6 r_H^6}\,.
\end{equation}
Then, by writing the event horizon as a function of the black hole charge, due to we have fixed $\Lambda$, and $M$, it is possible to show that the second derivative is positive when $-(1/12)^{1/4}<q<(1/12)^{1/4}$ or $-q_{ext}<q<q_{ext}$.

\begin{figure}
\begin{center}
\includegraphics[width=0.4\textwidth]{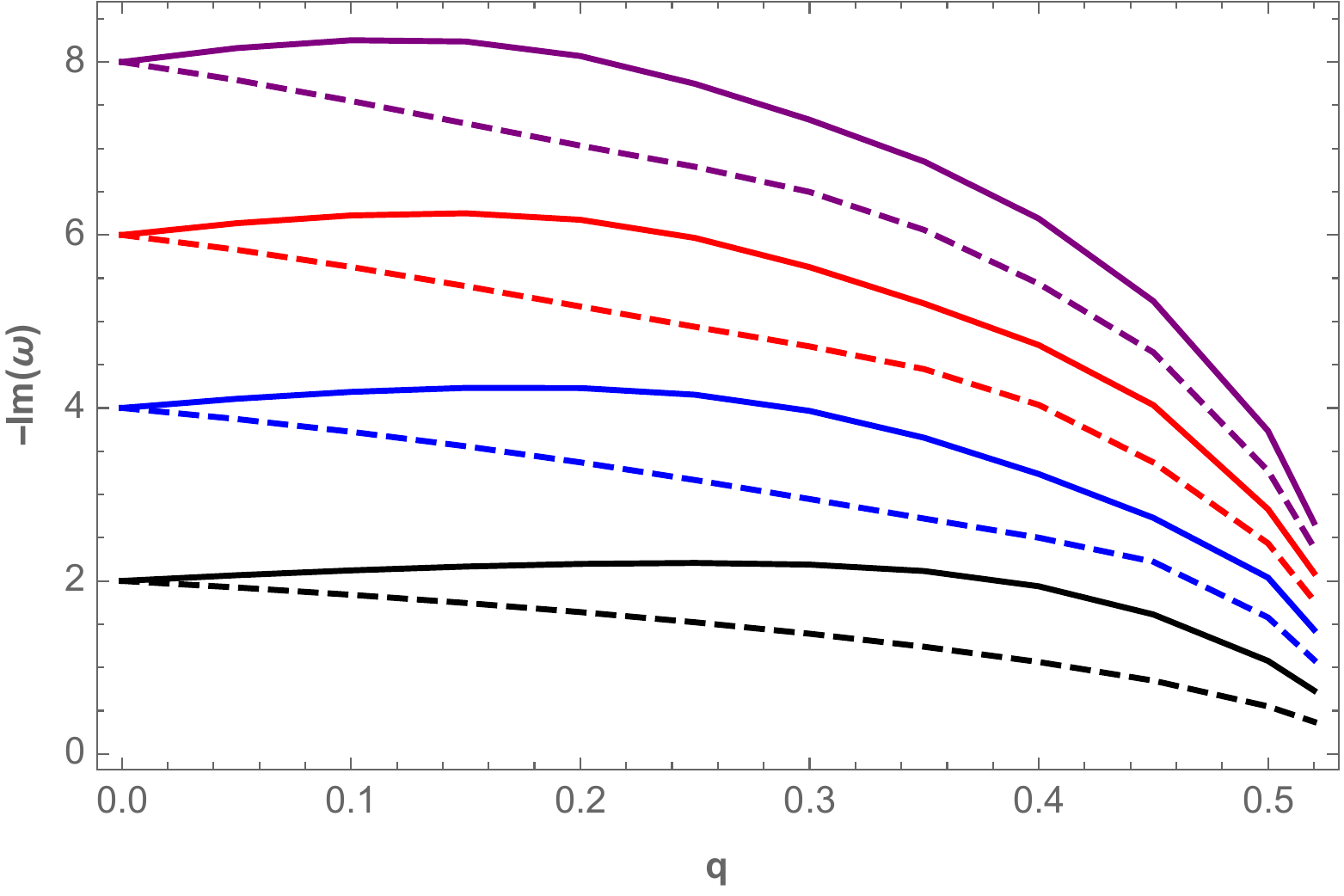}
\end{center}
\caption{QNFs for massless scalar fields in the background  of three-dimensional Coulomb like AdS black holes with $M=1$, $\Lambda=-1 $, $\kappa=0$, and different values of the overtone number $n$, and $q$.}
\label{Coulomb1}
\end{figure}

\begin {table}
\caption {QNFs for massless scalar fields in the background  of three-dimensional Coulomb like AdS black holes with $M=1$, $\Lambda = -1 $, $\kappa=0$, and different values of the overtone number $n$, and $q$.}
\label {TCoulomb1}\centering
\scalebox{0.8}{
\begin {tabular} { | c | c | c |}
\hline
${}$ & $q = 0.00$ & $q=0.05$   \\\hline
${}$ & $r_H = 1.000$ & $r_H=0.998$   \\\hline
$\omega(n=0)$ &
$-2.00000000 i$ &
$-1.92459850 i$  \\\hline
$\omega(n=1)$ &
$-4.00000000 i$ &
$-2.06592744 i$   \\\hline
$\omega(n=2)$ &
$-6.00000000 i$ &
$-3.87217357 i$  \\\hline
$\omega(n=3)$ &
$-8.00000000 i$ &
$-4.10579811 i$  \\\hline
${}$ & $q = 0.10$ & $q=0.15$   \\\hline
${}$ & $r_H = 0.993$ & $r_H=0.985$   \\\hline
$\omega(n=0)$ &
$- 1.83976265 i$ &
$- 1.74508060 i$  \\\hline
$\omega(n=1)$ &
$- 2.12181146 i$ &
$- 2.16632138 i$   \\\hline
$\omega(n=2)$ &
$- 3.72425650 i$ &
$- 3.55709399 i$  \\\hline
$\omega(n=3)$ &
$- 4.18567498 i$ &
$- 4.23206154 i$  \\\hline
${}$ & $q = 0.20$ & $q=0.25$   \\\hline
${}$ & $r_H = 0.972$ & $r_H=0.955$   \\\hline
$\omega(n=0)$ &
$- 1.63966533 i$ &
$- 1.52202672 i$  \\\hline
$\omega(n=1)$ &
$- 2.19682105 i$ &
$- 2.20806920 i$   \\\hline
$\omega(n=2)$ &
$- 3.37102742 i$ &
$- 3.16669614 i$  \\\hline
$\omega(n=3)$ &
$- 4.23017490 $ &
$- 4.15294463 i$  \\\hline
${}$ & $q = 0.30$ & $q=0.35$   \\\hline
${}$ & $r_H = 0.934$ & $r_H=0.905$   \\\hline
$\omega(n=0)$ &
$-1.38978111 i$ &
$-1.23902552 i$  \\\hline
$\omega(n=1)$ &
$-2.18892756 i$ &
$-2.11449442 i$   \\\hline
$\omega(n=2)$ &
$-2.94692496 i$ &
$-2.72085938 i$  \\\hline
$\omega(n=3)$ &
$-3.96584834 i$ &
$-3.65565875 i$  \\\hline
${}$ & $q = 0.40$ & $q=0.45$   \\\hline
${}$ & $r_H = 0.869$ & $r_H=0.819$   \\\hline
$\omega(n=0)$ &
$-1.06285206 i$ &
$-0.84699575 i$  \\\hline
$\omega(n=1)$ &
$-1.93793946 i$ &
$-1.61184343 i$   \\\hline
$\omega(n=2)$ &
$-2.49997056 i$ &
$-2.22164187 i$  \\\hline
$\omega(n=3)$ &
$-3.23575094 i$ &
$-2.72940071 i$  \\\hline
${}$ & $q = 0.50$ & $q=0.52$  \\\hline
${}$ & $r_H = 0.742$ & $r_H=0.692$   \\\hline
$\omega(n=0)$ &
$-0.55009852 i$ &
$-0.37170187 i$ \\\hline
$\omega(n=1)$ &
$-1.07523377 i$ &
$-0.73354719 i$ \\\hline
$\omega(n=2)$ &
$-1.57738460 i$  &
$-1.08895112 i$ \\\hline
$\omega(n=3)$ &
$-2.03801266 i$  &
$-1.43720563 i$ \\\hline
\end {tabular}}
\end{table}

\newpage

\item{\bf{Case $\kappa > 0$}}. The emergence of two branches disappears when the angular number is non null. In this case, the QNFs become complex in the majority of the cases, and we can observe that the effect of the inclusion of a like Coulomb field from non lineal electrodynamics is to decrease the frequency of the oscillation as well as the decay rate of the modes in comparison with the quasinormal frequencies of the static BTZ black hole, see Table \ref{TCoulomb2}.

\begin {table}[h]
\caption {The QNFs ($n=0,1$) for massless scalar fields in the background of  three-dimensional Coulomb like AdS black holes with $M=1$, $\Lambda = -1 $, and different values of the angular number $\kappa$ and $q$.}
\label {TCoulomb2}\centering
\scalebox{0.8}{
\begin {tabular} { | c | c | c |c |}
\hline
${}$ & $n$ & $q = 0.00$ & $q=0.05$   \\\hline
$\omega(\k=1)$ & $0$ &
$1.00000000 - 2.00000000 i$ &
$0.99602298 - 1.99640207 i$   \\\hline
$\omega(\k=1)$ & $1$ &
$1.00000000 - 4.00000000 i$ &
$0.98824152 - 3.99102850 i$  \\\hline
$\omega(\k=10)$ & $0$ &
$10.00000000 - 2.00000000 i$ &
$9.99875713 - 1.99876506 i$  \\\hline
$\omega(\k=10)$ & $1$ &
$10.00000000 - 4.00000000 i$ &
$9.99637929 - 3.99650203 i$  \\\hline
$\omega(\k=30)$ &$0$ &
$30.00000000 - 2.00000000 i$ &
$29.99928378 - 1.99928555 i$  \\\hline
$\omega(\k=30)$ & $1$ &
$30.00000000 - 4.00000000 i$ &
$29.99793092 - 3.99795569 i$  \\\hline

${}$ & $n$ & $q = 0.10$ & $q=0.15$   \\\hline
$\omega(\k=1)$ & $0$ &
$0.98388504 - 1.98552183 i$ &
$0.96292092 - 1.96709473 i$   \\\hline
$\omega(\k=1)$ & $1$ &
$0.95193811 - 3.96371161 i$ &
$0.88759003 - 3.91674137 i$  \\\hline
$\omega(\k=10)$ & $0$ &
$9.99500722 - 1.99505932 i$ &
$9.98868535 - 1.98888088 i$  \\\hline
$\omega(\k=10)$ & $1$ &
$9.98543115 - 3.98600214 i$ &
$9.96689163 - 3.96848693 i$  \\\hline
$\omega(\k=30)$ & $0$ &
$29.99712804 - 1.99714212 i$ &
$29.99351137 - 1.99356961 i$  \\\hline
$\omega(\k=30)$ & $1$ &
$29.99169525 - 3.99182217 i$ &
$29.98120677 - 3.98159861 i$  \\\hline

${}$ & $n$ & $q = 0.20$ & $q=0.25$   \\\hline
$\omega(\k=1)$ & $0$ &
$0.93185039 - 1.94066575 i$ &
$0.88841878 - 1.90558625 i$   \\\hline
$\omega(\k=1)$ & $1$ &
$0.78750745 - 3.84752359 i$ &
$0.63402688 - 3.75112754 i$  \\\hline
$\omega(\k=10)$ & $0$ &
$9.97967991 - 1.98022966 i$ &
$9.96782742 - 1.96911284 i$  \\\hline
$\omega(\k=10)$ & $1$ &
$9.94030081 - 3.94395057 i$ &
$9.90497217 - 3.91242570 i$  \\\hline
$\omega(\k=30)$ & $0$ &
$29.98839744 - 1.98856847 i$ &
$29.98173415 - 1.98214072 i$  \\\hline
$\omega(\k=30)$ & $1$ &
$29.96631831 - 3.96728694 i$ &
$29.94681672 - 3.94889750 i$  \\\hline

${}$ & $n$ & $q = 0.30$ & $q=0.35$   \\\hline
$\omega(\k=1)$ & $0$ &
$0.82856589 - 1.86110035 i$ &
$0.74421853 - 1.80696212 i$   \\\hline
$\omega(\k=1)$ & $1$ &
$0.36761948 - 3.61661927 i$ &
$-3.01785414 i$  \\\hline
$\omega(\k=10)$ & $0$ &
$9.95290546 - 1.95555416 i$ &
$9.93462415 - 1.93960939 i$  \\\hline
$\omega(\k=10)$ & $1$ &
$9.85995016 - 3.87404074 i$ &
$9.80396252 - 3.82912249 i$  \\\hline
$\omega(\k=30)$ & $0$ &
$29.97345230 - 1.97429145 i$ &
$29.96346405 - 1.96503123 i$  \\\hline
$\omega(\k=30)$ & $1$ &
$29.92241577 - 3.92645820 i$ &
$29.89274768 - 3.90002888 i$  \\\hline

${}$ & $n$ & $q = 0.40$ & $q=0.45$   \\\hline
$\omega(\k=1)$ & $0$ &
$0.61560316 - 1.74821524 i$ &
$-1.62683274 i$   \\\hline
$\omega(\k=1)$ & $1$ &
$-2.44345944 i$ &
$0.42137792 - 1.78237794 i$  \\\hline
$\omega(\k=10)$ & $0$ &
$9.91261854 - 1.92139101 i$ &
$9.88644683 - 1.90110639 i$  \\\hline
$\omega(\k=10)$ & $1$ &
$9.73539151 - 3.77837308 i$ &
$9.65232994 - 3.72315020 i$  \\\hline
$\omega(\k=30)$ & $0$ &
$29.9516611 - 1.95437946 i$ &
$29.9379131 - 1.94236919 i$  \\\hline
$\omega(\k=30)$ & $1$ &
$29.8573542 - 3.86972297 i$ &
$29.8156803 - 3.83573962 i$  \\\hline

${}$ & $n$ & $q = 0.50$ & $q=0.52$  \\\hline
$\omega(\k=1)$ & $0$ &
$-0.89226643 i$ &
$-0.59057899 i$   \\\hline
$\omega(\k=1)$ & $1$ &
$-1.48084248 i$ &
$-0.95426553 i$  \\\hline
$\omega(\k=10)$ & $0$ &
$9.85560581 - 1.87911189 i$ &
$9.84184537 - 1.86995434 i$  \\\hline%
$\omega(\k=10)$ & $1$ &
$9.55286027 - 3.66582543 i$ &
$9.50815121 - 3.64307665 i$  \\\hline
$\omega(\k=30)$ & $0$ &
$29.92206613 - 1.92905381 i $ &
$29.91510074 - 1.92337954 i $  \\\hline
$\omega(\k=30)$ & $1$ &
$29.76707552 - 3.79840931 i$ &
$29.74553186 - 3.78264900 i$  \\\hline
\end{tabular}}
\end{table}

\newpage

Here, by considering $M=1$, $\Lambda=-1$, $q=0$, and $\kappa \neq 0$, the second derivative is $-6\kappa^2$, so it is always negative, and it is associated with complex QNFs. 
However, for a charge black hole, we can see from Table \ref{TCoulomb2} the existence of purely and complex QNFs for $q>0.35$. So, we analyze the existence of a change of concavity for these values of the charge. For instance, for $q=0.35$, we obtain that the second derivative is negative if $\kappa > 0.556$; for $q=0.40$, $\kappa > 0.676$; for $q=0.45$, $\kappa > 0.857$; for $q=0.50$, $\kappa >1.470$; and for $q=0.52$ always is positive. Due to the existence of purely imaginary and complex QNFs for a same value of black hole charge, one could think in the existence of two families of modes, in order to explain this behaviour or alternatively that the complex QNFs become purely imaginary. In order to elucidate this, we plot in Fig. \ref{families} different overtone numbers for $\kappa=1$. We can observe that there are complex QNFs for small values of the charge that then become in two branches of imaginary QNFs when the black hole charge increases; thereby,  for small values of the black hole charge the complex QNFs are dominant, while that for bigger values of the black hole charge the purely imaginary QNFs are dominant. Also, it is possible to observe that the value of the charge for which occurs decreases when the overtone number increases.

\begin{figure}[h]
\begin{center}
\includegraphics[width=0.45\textwidth]{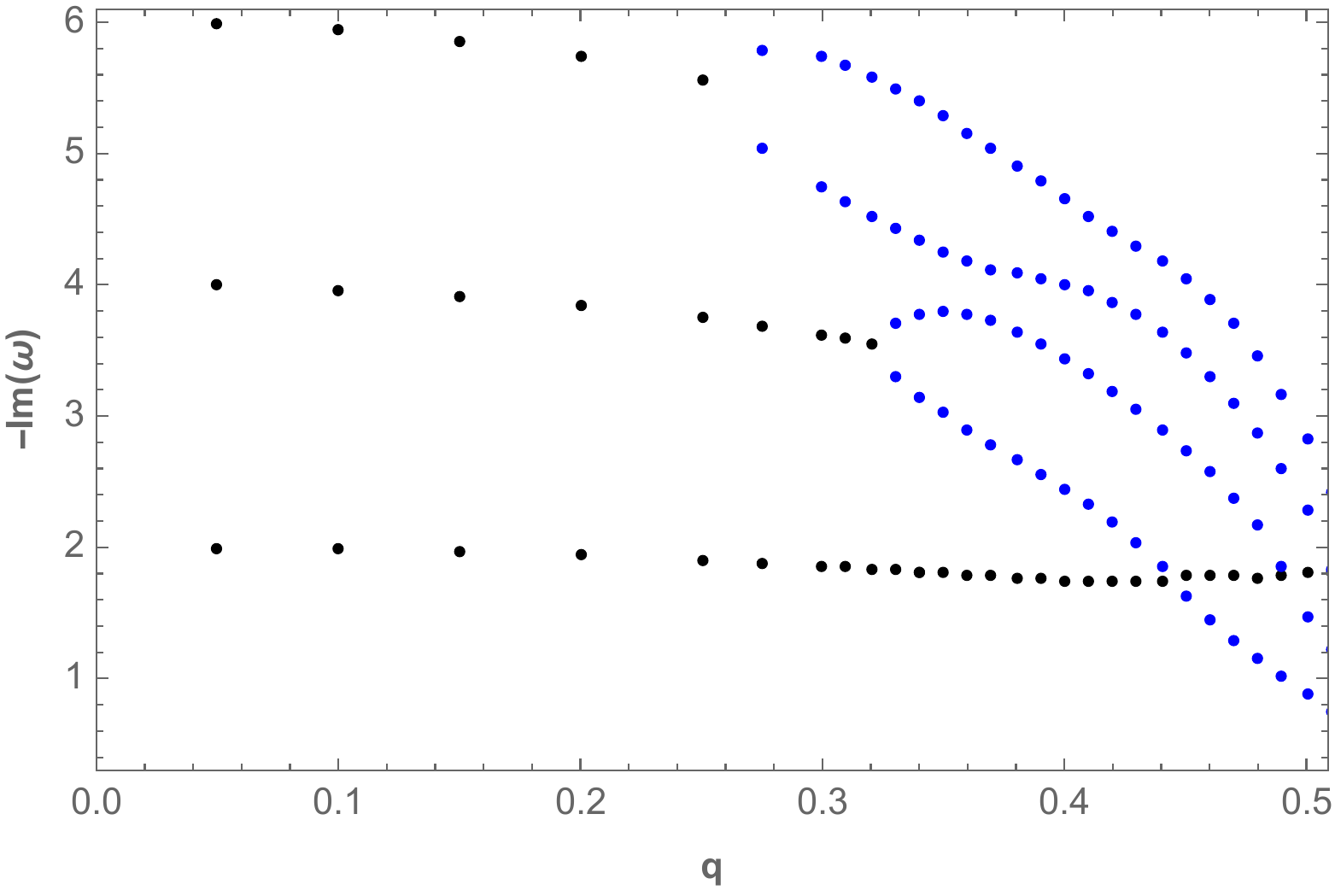}
\end{center}
\caption{QNFs for massless scalar fields in the background  of three-dimensional Coulomb like AdS black holes with $M=1$, $\Lambda=-1 $, $\kappa=1$, and different values of the overtone number $n$, and $q$. Black points correspond to complex QNFs, while that blue points correspond to purely imaginary QNFs. Here, we show $n=0,1,2$, for small black hole charge and $n=0,1,2,3,4$, for bigger black hole charge.}
\label{families}
\end{figure}

In the following we consider massive scalar field in order to analyze the effect of the scalar field mass in the QNFs.


\end{itemize}

\subsubsection{Massive scalar fields}

\begin{itemize}

\item  
{\bf{Case $\kappa=0$.}} To analyze the propagation of massive scalar fields we show the behaviour of the QNFs for massive scalar fields for fixed values of black hole mass, cosmological constant, black hole charge, and a vanishing angular number, and different values of the overtone number and scalar field mass, see Table \ref{TCoulomb3}, where we can observe that when the scalar field mass increases the decay rate increases. Here, the emergence of two branches is present, also note that the separation between $-Im(\omega)$ observed in Fig. \ref{Coulomb1} for $q=0.25$, remain almost constant when we added a scalar field mass in the interval for the scalar field mass [0, 0.30], see Fig. \ref{Coulomb3}.
Also, for $M=1$, $\Lambda=-1$, $\kappa=0$, and $q=0.25$, the event horizon is located at $r_H=0.955$, and the second derivative of the potential is $0.820+2.191 m^2$
which is always positive; thereby there is not a change of concavity for the effective potential.\\

\begin{figure}[h!]
\begin{center}
\includegraphics[width=0.4\textwidth]{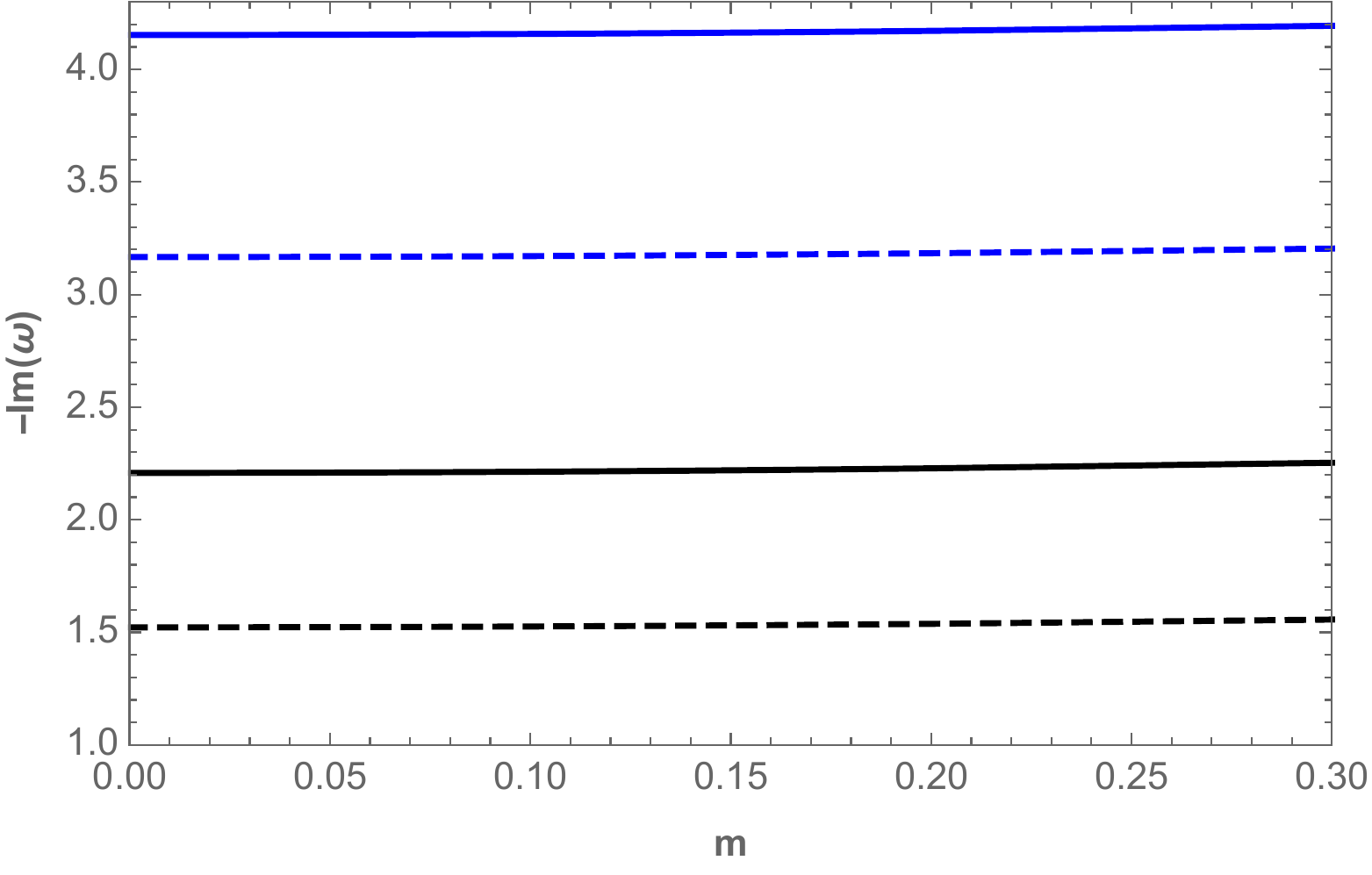}
\end{center}
\caption{QNFs for massive scalar fields in the background  of three-dimensional Coulomb like AdS black holes with $M=1$, $q=0.25$, $\Lambda=-1 $, $\kappa=0$, and different values of the overtone number $n$, and $m$.}
\label{Coulomb3}
\end{figure}

\begin {table}
\caption {QNFs for massive scalar fields in the background of three-dimensional Coulomb like AdS black holes with $M=1$, $\Lambda = -1 $, $\kappa=0$, and different values of the overtone number $n$, and $q=0.25$.}
\label {TCoulomb3}\centering
\scalebox{0.8}{
\begin {tabular} { | c | c | c |}
\hline
${}$ & $m = 0.00$ & $m=0.02$   \\\hline
$\omega(n=0)$ &
$-1.52202672 i$ &
$-1.52218426 i$  \\\hline
$\omega(n=1)$ &
$-2.20806920 i$ &
$-2.20827406 i$   \\\hline
$\omega(n=2)$ &
$-3.16669614 i$ &
$-3.16686707 i$  \\\hline
$\omega(n=3)$ &
$-4.15294463 i$ &
$-4.15313052 i$  \\\hline
${}$ & $m = 0.04$ & $m=0.06$   \\\hline
$\omega(n=0)$ &
$-1.52265669 i$ &
$-1.52344346 i$  \\\hline
$\omega(n=1)$ &
$-2.20888838 i$ &
$-2.20991139 i$   \\\hline
$\omega(n=2)$ &
$-3.16737966 i$ &
$-3.16823330 i$  \\\hline
$\omega(n=3)$ &
$-4.15368798 i$ &
$-4.15461631 i$  \\\hline
${}$ & $m = 0.08$ & $m=0.10$   \\\hline
$\omega(n=0)$ &
$-1.52454364 i$ &
$-1.52595592 i$  \\\hline
$\omega(n=1)$ &
$-2.21134178 i$ &
$-2.21317775 i$   \\\hline
$\omega(n=2)$ &
$-3.16942704 i$ &
$-3.17095948 i$  \\\hline
$\omega(n=3)$ &
$-4.15591436 i$ &
$-4.15758054 i$  \\\hline
${}$ & $m = 0.12$ & $m=0.14$   \\\hline
$\omega(n=0)$ &
$-1.52767867 i$ &
$-1.52970989 i$  \\\hline
$\omega(n=1)$ &
$-2.21541700 i$ &
$-2.21805676 i$   \\\hline
$\omega(n=2)$ &
$-3.17282890 i$ &
$-3.17503318 i$  \\\hline
$\omega(n=3)$ &
$-4.15961279 i$ &
$-4.16200867 i$  \\\hline
${}$ & $m = 0.16$ & $m=0.18$   \\\hline
$\omega(n=0)$ &
$-1.53204723 i$ &
$-1.53468806 i$  \\\hline
$\omega(n=1)$ &
$-2.22109377 i$ &
$-2.22452433 i$   \\\hline
$\omega(n=2)$ &
$-3.17756985 i$ &
$-3.18043612 i$  \\\hline
$\omega(n=3)$ &
$-4.16476527 i$ &
$-4.16787933 i$  \\\hline
${}$ & $m = 0.20$ & $m=0.30$  \\\hline
$\omega(n=0)$ &
$-1.53762941 i$ &
$-1.55672085 i$ \\\hline
$\omega(n=1)$ &
$-2.22834431 i$ &
$-2.25311416 i$ \\\hline
$\omega(n=2)$ &
$-3.18362884 i$  &
$-3.20435931 i$ \\\hline
$\omega(n=3)$ &
$-4.17134719 i$  &
$-4.19384134 i$ \\\hline
\end{tabular}}
\end{table}

\newpage

\item  
{\bf{Case $\kappa > 0$.}} As in the massless scalar field case, the emergence of two branches 
disappears when angular number is not null. In this case, the QNFs become complex and as we have previously discussed it depends on the value of black hole charge and the angular number, as well as, on the overtone number. For $M=1$, $q=0.25$, $\Lambda=-1$, and $m=0.30$ the second derivative of the effective potential is negative if $\kappa > 0.413$ (for massless case $\kappa > 0.371$), by increasing the value of $\kappa$. Also, we can observe that when the scalar field mass increases the frequency of the oscillation decreases, and the decay rate increases, see Table \ref{TCoulomb4}, where, we have considered the fundamental QNFs ($n=0$).
On the other hand, note that the longest-lived modes are the ones with smallest angular number, contrary to Schwarzschild-AdS and Reissner-Nordstr\"om-AdS spacetimes where appear an anomalous behaviour of the decay rate, i.e the longest-lived modes are the ones with higher angular number for small values of the scalar field mass \cite{Aragon:2020tvq,Fontana:2020syy}. In this context one could say the anomalous behavior of the decay rate depend on dimension of the spacetime, being it possible for spacetime with dimensions greater than three.

\begin {table}
\caption {The fundamental QNFs ($n=0$) for massive scalar fields in the background of three-dimensional Coulomb like AdS black holes with $M=1$, $\Lambda = -1 $, and different values of $\kappa$ and $q=0.25$.}
\label {TCoulomb4}\centering
\scalebox{0.8}{
\begin {tabular} { | c | c | c |}
\hline
${}$ & $m = 0.00$ & $m=0.02$   \\\hline
$\omega(\k=1)$ &
$0.88841878 - 1.90558625 i$ &
$0.88840192 - 1.90577346 i$   \\\hline
$\omega(\k=10)$ &
$9.96782742 - 1.96911284 i$ &
$9.96782298 - 1.96930864 i$  \\\hline
$\omega(\k=30)$ &
$29.98173415 - 1.98214072 i$ &
$29.98173165 - 1.98233827 i$  \\\hline
${}$ & $m = 0.04$ & $m=0.06$   \\\hline
$\omega(\k=1)$ &
$0.88835135 - 1.90633485 i$ &
$0.88826711 - 1.90726975 i$   \\\hline
$\omega(\k=10)$ &
$9.96780965 - 1.96989580 i$ &
$9.96778745 - 1.97087362 i$  \\\hline
$\omega(\k=30)$ &
$29.98172414 - 1.98293070 i$ &
$29.98171163 - 1.98391729 i$  \\\hline
${}$ & $m = 0.08$ & $m=0.10$   \\\hline
$\omega(\k=1)$ &
$0.88814928 - 1.90857702 i$ &
$0.88799795 - 1.91025508 i $   \\\hline
$\omega(\k=10)$ &
$9.96775639 - 1.97224093 i$ &
$9.96771651 - 1.97399610 i$  \\\hline
$\omega(\k=30)$ &
$29.98169414 - 1.98529686 i$ &
$29.98167168 - 1.98706776 i$  \\\hline
${}$ & $m = 0.12$ & $m=0.14$   \\\hline
$\omega(\k=1)$ &
$0.88781325 - 1.91230191 i$ &
$0.88759533 - 1.91471508 i$   \\\hline
$\omega(\k=10)$ &
$9.96766784 - 1.97613703 i$ &
$9.96761043 - 1.97866123 i$  \\\hline
$\omega(\k=30)$ &
$29.98164427 - 1.98922791 i$ &
$29.98161193 - 1.99177477 i$  \\\hline
${}$ & $m = 0.16$ & $m=0.18$   \\\hline
$\omega(\k=1)$ &
$0.88734437 - 1.91749175 i$ &
$0.88706057 - 1.92062868 i$   \\\hline
$\omega(\k=10)$ &
$9.96754433 - 1.98156574 i$ &
$9.96746959 - 1.98484722 i$  \\\hline
$\omega(\k=30)$ &
$29.98157470 - 1.99470538 i$ &
$29.98153261 - 1.99801635 i$  \\\hline
${}$ & $m = 0.20$ & $m=0.30$  \\\hline
$\omega(\k=1)$ &
$0.88674415 - 1.92412224 i$ &
$0.88468224 - 1.94678943 i$   \\\hline
$\omega(\k=10)$ &
$9.96738628 - 1.98850193 i$ &
$9.96684391 - 2.01221883 i$  \\\hline
$\omega(\k=30)$ &
$29.98148569 - 2.00170395 i$ &
$29.98118026 - 2.02563487 i$  \\\hline
\end {tabular}}
\end{table}

\end{itemize}


\subsubsection{Near extremal}

\begin{itemize}

\item {\bf{Case $\kappa=0$.}} Now, we analyze the near extremal case, for massless scalar fields and a null angular number, we can observe that the separation observed decreases when the charge of the black hole tends to the $q_{ext}$, see Fig. \ref{Extremal}. The numerical values are in Table \ref{Extremal}.

\begin{figure}[h!]
\begin{center}
\includegraphics[width=0.4\textwidth]{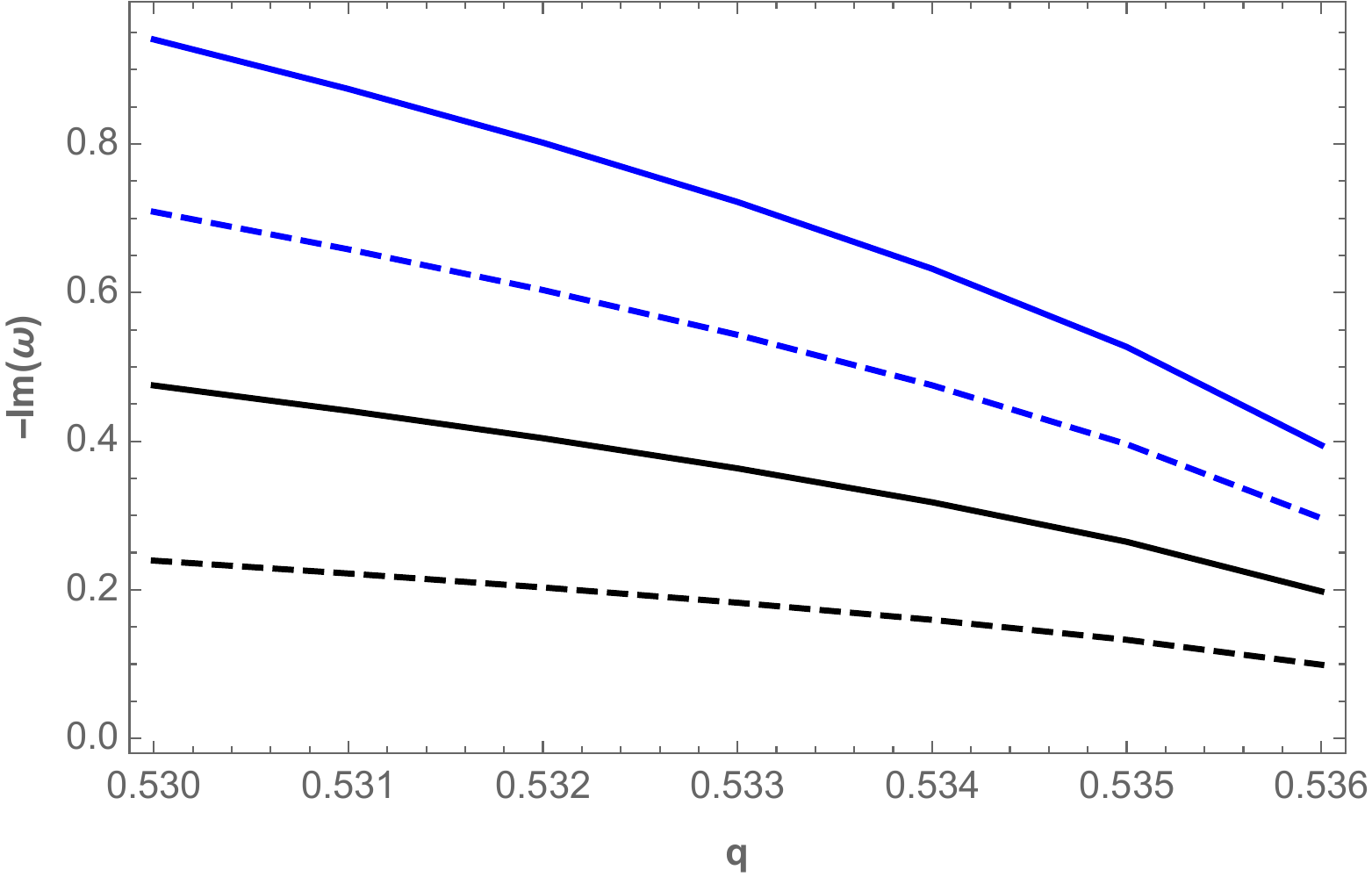}
\end{center}
\caption{QNFs for massless scalar fields in the background  of a near extremal three-dimensional Coulomb like AdS black holes with $M=1$, $\Lambda=-1 $, $\kappa=0$, and different values of the overtone number $n$, and $q$.}
\label{Extremal}
\end{figure}

\begin {table}[h]
\caption {QNFs for massless scalar fields in the background  of near extremal three-dimensional Coulomb like AdS black holes with $M=1$, $\Lambda = -1 $, $\kappa=0$, and different values of the overtone number $n$, and $q$; $\Delta_r=r_+-r_-$.}
\label {TCoulomb5}\centering
\scalebox{0.8}{
\begin {tabular} { | c | c | c |}
\hline
${}$ & $q = 0.530$ & $q=0.531$   \\\hline
${}$ & $\Delta_r=0.1549$ & $\Delta_r=0.1439$ \\\hline
$\omega(n=0)$ &
$-0.23929763 i$ &
$-0.22197311 i$  \\\hline
$\omega(n=1)$ &
$-0.47499114 i$ &
$-0.44091025 i$  \\\hline
$\omega(n=2)$ &
$-0.70861507 i$ &
$-0.65811723 i$  \\\hline
$\omega(n=3)$ &
$-0.94050627 i$ &
$-0.87391400 i$  \\\hline
${}$ & $q = 0.532$ & $q=0.533$   \\\hline
${}$ & $\Delta_r=0.1320$ & $\Delta_r=0.1189$ \\\hline
$\omega(n=0)$ &
$-0.20324315 i$ &
$-0.18268786 i$  \\\hline
$\omega(n=1)$ &
$-0.40400482 i$ &
$-0.36343308 i$  \\\hline
$\omega(n=2)$ &
$-0.60335990 i$ &
$-0.54307959 i$  \\\hline
$\omega(n=3)$ &
$-0.80159793 i$ &
$-0.72187489 i$  \\\hline
${}$ & $q = 0.534$ & $q=0.535$   \\\hline
${}$ & $\Delta_r=0.1042$ & $\Delta_r=0.0869$ \\\hline
$\omega(n=0)$ &
$-0.15962240 i$ &
$-0.13277855 i$  \\\hline
$\omega(n=1)$ &
$-0.31782221 i$ &
$-0.26463047 i$  \\\hline
$\omega(n=2)$ &
$-0.47521540 i$ &
$-0.39595133 i$  \\\hline
$\omega(n=3)$ &
$-0.63199769 i$ &
$-0.52687750 i$  \\\hline
${}$ & $q = 0.536$ & ${}$   \\\hline
${}$ & $\Delta_r=0.0652$ & ${}$ \\\hline
$\omega(n=0)$ &
$-0.09921077 i$ &
${}$  \\\hline
$\omega(n=1)$ &
$-0.19795596 i$ &
${}$  \\\hline
$\omega(n=2)$ &
$-0.29642676 i$ &
${}$  \\\hline
$\omega(n=3)$ &
$-0.39469568 i$ &
${}$  \\\hline
\end {tabular}}
\end{table}

\newpage

\item{\bf{Case $\kappa>0$.}} Firstly, notice that the number of  fundamental QNFs that are purely imaginary increases when the black hole charge increases, and it also depend on the angular number, for instance, for $q=0.45$ there is only one purely imaginary fundamental QNF, and it is for $\kappa=1$, while that for $q=0.53$ the fundamental QNFs are purely imaginary for $\kappa=1,2,3,4,5$, see Table \ref{Coulomb6}. For, the near extremal case, see Table \ref{TCoulomb7}, we can observe that for $q=0.532$ and $\kappa=1,2,...,8$ the fundamental QNFs are purely imaginary, while that for $q=0.536$, the fundamental QNFs are purely imaginary for $\kappa=1,2,...,18$, by increasing the set of values for the angular number when the black hole charge increases.

\begin {table}
\caption {Fundamental modes ($n =
      0 $) for massless scalar fields,  in the background of three-dimensional Coulomb like AdS black holes with $M=1$, $\Lambda = -1 $, and different values of $\kappa$, and $q$.}
\label {Coulomb6}\centering
\scalebox{0.8}{
\begin {tabular} { | c | c | c |}
\hline
${}$ & $q = 0.44$ & $q=0.45$  \\\hline
$\omega(\kappa=1)$ &
$0.44652734 - 1.74931395 i$ &
$-1.62683274 i$  \\\hline
$\omega(\kappa=2)$ &
$1.70978492 - 1.79697888 i$ &
$1.69205258 - 1.79017800 i$  \\\hline
$\omega(\kappa=3)$ &
$2.77903855 - 1.83140317 i$ &
$2.76633073 - 1.82476709 i$  \\\hline
$\omega(\kappa=4)$ &
$3.81547923 - 1.85274359 i$ &
$3.80520685 - 1.84663880 i$  \\\hline
$\omega(\kappa=5)$ &
$4.83877163 - 1.86759237 i$ &
$4.82998483 - 1.86195874 i$  \\\hline
$\omega(\kappa=10)$ &
$9.89203800 - 1.90531417 i$ &
$9.88644683 - 1.90110639 i$  \\\hline
$\omega(\kappa=20)$ &
$19.92638377 - 1.93263473 i$ &
$19.92270386 - 1.92958602 i$  \\\hline
$\omega(\kappa=30)$ &
$29.94082535 - 1.94487772 i$ &
$29.93791309 - 1.94236919 i$  \\\hline
${}$ & $q = 0.46$ & $q=0.47$  \\\hline
$\omega(\kappa=1)$ &
$-1.43830233 i$ &
$-1.28749240 i$  \\\hline
$\omega(\kappa=2)$ &
$1.67366148 - 1.78379307 i$ &
$1.65466952 - 1.77787728 i$  \\\hline
$\omega(\kappa=3)$ &
$2.75311936 - 1.81821475 i$ &
$2.73940580 - 1.81178030 i$  \\\hline
$\omega(\kappa=4)$ &
$3.79453892 - 1.84052897 i$ &
$2.73940580 - 1.81178030 i$  \\\hline
$\omega(\kappa=5)$ &
$4.82087040 - 1.85628553 i$ &
$4.81142183 - 1.85058611 i$  \\\hline
$\omega(\kappa=10)$ &
$9.88066894 - 1.89682954 i$ &
$9.87470011 - 1.89248735 i$  \\\hline
$\omega(\kappa=20)$ &
$19.91891266 - 1.92647673 i$ &
$19.91500808 - 1.92330796 i$  \\\hline
$\omega(\kappa=30)$ &
$29.93491695 - 1.93980836 i$ &
$29.93183559 - 1.93719578 i$  \\\hline
${}$ & $q = 0.48$ & $q=0.49$  \\\hline
$\omega(\kappa=1)$ &
$-1.15298727 i$ &
$-1.02365438 i$  \\\hline
$\omega(\kappa=2)$ &
$1.63513833 - 1.77246638 i$ &
$-1.69893649 i$  \\\hline
$\omega(\kappa=3)$ &
$2.72519595 - 1.80549826 i$ &
$2.71050030 - 1.79940261 i$  \\\hline
$\omega(\kappa=4)$ &
$3.77199576 - 1.82837649 i$ &
$3.76011514 - 1.82237796 i$  \\\hline
$\omega(\kappa=5)$ &
$4.80163345 - 1.84487497 i$ &
$4.79150061 - 1.83916761 i$  \\\hline
$\omega(\kappa=10)$ &
$9.86853610 - 1.88808394 i$ &
$9.86217272 - 1.88362382 i$  \\\hline
$\omega(\kappa=20)$ &
$19.91098799 - 1.92008090 i$ &
$19.90685020 - 1.91679688 i$  \\\hline
$\omega(\kappa=30)$ &
$29.78739437 - 3.81371473 i$ &
$29.92541161 - 1.93181783 i$  \\\hline
${}$ & $q = 0.50$ & $q=0.51$  \\\hline
$\omega(\kappa=1)$ &
$-0.89226643 i$ &
$-0.75164892 i$  \\\hline
$\omega(\kappa=2)$ &
$-1.47136414 i$ &
$-1.22836213 i$  \\\hline
$\omega(\kappa=3)$ &
$2.69533369 - 1.79352579 i$ &
$-1.70195256 i$  \\\hline
$\omega(\kappa=4)$ &
$3.74782868 - 1.81646204 i$ &
$3.73513963 - 1.81065229 i$  \\\hline
$\omega(\kappa=5)$ &
$4.78101993 - 1.83348047 i$ &
$4.77018941 - 1.82783077 i$  \\\hline
$\omega(\kappa=10)$ &
$9.85560581 - 1.87911189 i$ &
$9.84883133 - 1.87455348 i$  \\\hline
$\omega(\kappa=20)$ &
$19.90259249 - 1.91345732 i$ &
$19.89821260 - 1.91006379 i$  \\\hline
$\omega(\kappa=30)$ &
$29.92206613 - 1.92905381 i$ &
$29.91862968 - 1.92624077 i$  \\\hline
${}$ & $q = 0.52$ & $q=0.53$  \\\hline
$\omega(\kappa=1)$ &
$-0.59057899 i$ &
$-0.37926211 i$  \\\hline
$\omega(\kappa=2)$ &
$-0.95488654 i$ &
$-0.60604255 i$  \\\hline
$\omega(\kappa=3)$ &
$-1.32223623 i$ &
$-0.83852483 i$  \\\hline
$\omega(\kappa=4)$ &
$-1.69271449 i$ &
$-1.07314674 i$  \\\hline
$\omega(\kappa=5)$ &
$4.75900863 - 1.82223633 i$ &
$-1.30866231 i$  \\\hline
$\omega(\kappa=10)$ &
$9.84184537 - 1.86995434 i$ &
$9.83464423 - 1.86532067 i$  \\\hline
$\omega(\kappa=20)$ &
$19.89370824 - 1.90661798 i$ &
$19.88907707 - 1.90312174 i$  \\\hline
$\omega(\kappa=30)$ &
$29.91510074 - 1.92337954 i$ &
$29.91147775 - 1.92047102 i$  \\\hline
\end {tabular}}
\end{table}

\newpage

\begin {table}[h]
\caption {The fundamental QNFs for massless scalar fields in the background of near extremal three-dimensional Coulomb like AdS black holes with $M=1$, $\Lambda = -1 $, and different values of $q$, and $\kappa$; $\Delta_r=r_+-r_-$.}
\label {TCoulomb7}\centering
\scalebox{0.8}{
\begin {tabular} { | c | c | c |}
\hline
${}$ & $q = 0.532$ & $q=0.536$   \\\hline
${}$ & $\Delta_r=0.1320$ & $\Delta_r=0.0652$ \\\hline
$\omega(\kappa=1)$ &
$-0.32240963 i$ &
$-0.15841160 i$  \\\hline
$\omega(\kappa=2)$ &
$-0.51392662 i$ &
$-0.25122736 i$  \\\hline
$\omega(\kappa=3)$ &
$-0.71093292 i$ &
$-0.34737912 i$  \\\hline
$\omega(\kappa=4)$ &
$-0.90979395 i$ &
$-0.44448563 i$  \\\hline
$\omega(\kappa=5)$ &
$-1.10942753 i$ &
$-0.54198520 i$  \\\hline
$\omega(\kappa=6)$ &
$-1.30945459 i$ &
$-0.63968366 i$  \\\hline
$\omega(\kappa=7)$ &
$-1.50970875 i$ &
$-0.73749650 i$  \\\hline
$\omega(\kappa=8)$ &
$-1.71010571 i$ &
$-0.83538109 i$  \\\hline
$\omega(\kappa=9)$ &
$8.82239581 - 1.85763733 i$ &
$-0.93331363 i$  \\\hline
$\omega(\kappa=10)$ &
$9.83317786 - 1.86439037 i$ &
$-1.03127980 i$  \\\hline
$\omega(\kappa=11)$ &
$10.84229142 - 1.87027541 i$ &
$-1.12927045 i$  \\\hline
$\omega(\kappa=12)$ &
$11.85011993 - 1.87546181 i$ &
$-1.22727947 i$  \\\hline
$\omega(\kappa=13)$ &
$12.85693486 - 1.88007651 i$ &
$-1.32530265 i$  \\\hline
$\omega(\kappa=14)$ &
$13.86293427 - 1.88421664 i$ &
$-1.42333694 i$  \\\hline
$\omega(\kappa=15)$ &
$14.86826640 - 1.88795785 i$ &
$-1.52138014 i$  \\\hline
$\omega(\kappa=16)$ &
$15.87304462 - 1.89136002 i$ &
$-1.61943057 i$  \\\hline
$\omega(\kappa=17)$ &
$16.87735724 - 1.89447122 i$ &
$-1.71748697 i$  \\\hline
$\omega(\kappa=18$ &
$17.88127421 - 1.89733053 i$ &
$-1.81554833 i$  \\\hline
$\omega(\kappa=19)$ &
$18.88485170 - 1.89997010 i$ &
$18.88289136 - 1.89852224 i$  \\\hline
$\omega(\kappa=20)$ &
$19.88813542 - 1.90241661 i$ &
$19.88623653 - 1.90100060 i$  \\\hline
\end {tabular}}
\end{table}

\end{itemize}

\newpage

\section{Final remarks}
\label{conclusion}

In this work we studied the propagation of scalar fields in the background of $2+1$-dimensional like Coulomb AdS black holes, and we showed that such propagation is stable under Dirichlet boundary conditions. Then, we solved the Klein-Gordon equation by using the pseudospectral Chevyshev method, and we found the quasinormal frequencies. Mainly, we found that the quasinormal frequencies are purely imaginary when the angular number is null. Also, we showed that the effect of the inclusion of a like Coulomb field from non lineal electrodynamics is the emergence of two branches of quasinormal frequencies by comparing with quasinormal frequencies of the static BTZ spacetime, when the angular number is null. Also, we showed that the existence of purely imaginary frequencies is associated with a positive value of the second derivative of the potential respect to $r$, evaluated at the event horizon. Then, when the angular number is not null, the QNFs are complex for small values of the black hole charge, that is associated with a negative value of the second derivative of the potential at the event horizon; however when the black hole charge increases, we have shown that it complex quasinormal frequency becomes in two branches of imaginary QNFs; thereby for small values of the black hole charge the complex QNFs are dominant, while that for bigger values of the black hole charge the purely imaginary QNFs are dominant. Also, we showed that the value of the charge for which occurs the change from complex to purely imaginary quasinormal frequencies decreases when the overtone number increases. It is worth mentioning that it would be interesting to analyze the superradiant instability of this  black hole for charged massive scalar field, as well as, the quasinormal modes which we left for a future work. \\

\acknowledgments
This work is partially supported by ANID Chile through FONDECYT Grant No 1170279 (J. S.).

\appendix

\end{document}